# Real-time Optical Network Sensing Control Plane Enabled by a Novel Sub-μs Response Time Fibre Sensing Control Device

Mijail Szczerban[(1)], Mikael Mazur[(1)], Lauren Dallachiesa[(1)], Haïk Mardoyan[(2)], Sarvesh Bidkar[(1)], Roland Ryf[(1)], Jesse Simsarian[(1)]

[(1)] Nokia Bell Labs, Murray Hill, NJ, USA, mijail.szczerban_gonzalez@nokia-bell-labs.com
[(2)] Nokia Bell Labs, Massy, France

**Abstract** *We propose and implement a novel fibre sensing control device and associated sensing control plane that effectively controls backscatter and polarization-based fibre sensing. We experimentally demonstrate in a fibre network that this device and associated control plane can achieve sub-μs response time.*

**Optical Fibre Sensing Network**

Network sensing is recognized as a key element for future networks and as a fundamental pillar for 6G [1,2]. Optical fibre sensing can detect events that impact optical transmission so that the network can quickly react. Backscatter light sensing (BLS) systems such as optical time domain reflectometry (OTDR) and distributed acoustic sensing (DAS) [2-5], provide information about the fibre network infrastructure, and mechanical events occurring in the proximity of fibre networks. Tracking the state of polarization (SoP) of optical communication signals also allows the detection of events in and around the fibre [6-8]. Fibre sensing has multiple networking applications that can increase network efficiency, reduce design margins, and localize fibre breaks; as well as environmental sensing applications such as, distributed temperature and acoustic sensing; wildfire, earthquake, and intrusion detection [2-8].

However, such powerful tools can become a security concern if used by ill-intentioned actors. Infrastructure details can be obtained using OTDRs and conversations in the proximity of exposed fibre can be detected using DAS. Therefore, providing control over where, when, and how, fibre sensing occurs is critical for future optical networks. In this work we propose and implement for the first time a fibre sensing control device (FSCD) and the associated sensing control plane that provides network owners and end users control over fibre sensing activities.

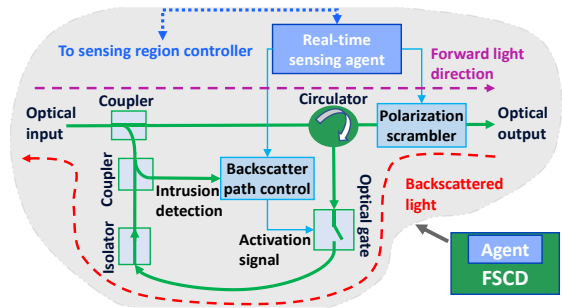

**Fig. 2:** Backscatter and state-of-polarization fibre sensing control device (FSCD)

The FSCD can be placed in strategic locations in the fibre infrastructure creating demarcation points and sensing regions where sensing activities are independently controlled through the sensing control plane as shown in Fig. 1.

**Optical Fibre Sensing Control Device**

The FSCD device we propose in this work controls both SoP sensing (SoPS) and backscatter-based sensing such as OTDR and DAS. Fig. 2 shows a schematic representation of a reconfigurable FSCD, including the real-time sensing agent. The first element from the left side of the figure is an optical coupler that merges the forward and backpropagating paths. The forward propagation direction is shown by the purple dashed arrow on the top of the figure. Forward-propagating light passes the coupler through the top arm and reaches the optical circulator that directs the light to the optical output port. A polarization scrambler is placed before the output port, affecting both propagation directions. To limit SoPS, the polarization scrambler is activated, varying the state of polarization to obscure certain polarization transient events (depending on the frequency of scrambling) that could be detected through SoP tracking at the receiver. The red arrow at the bottom of Fig. 2 represents the backscattered light path. Light entering from the output port (backscatter) is directed by the optical circulator to a secondary path which is controlled by an optical gate. An optical isolator is used after the optical gate to prevent light in the forward direction from reaching the gate. The real-time sensing agent provides the fast electronic signals to control SoPS and BLS, executing sensing policies defined by the sensing control plane.

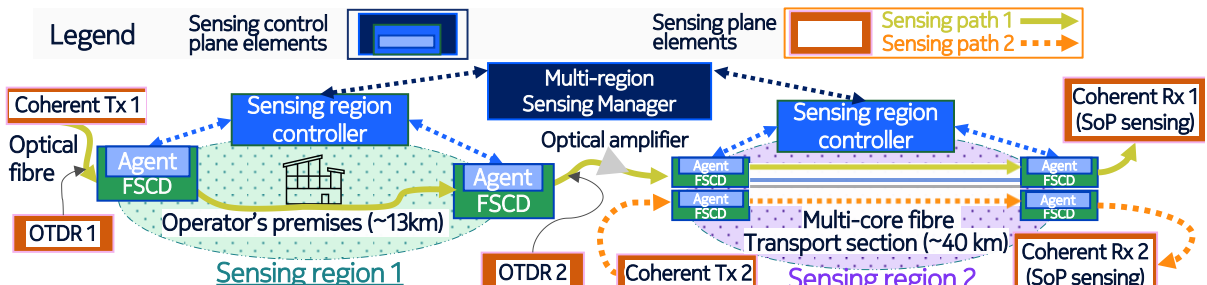

**Fig. 1:** Sensing network control conceptual and experimental scheme with FSCDs demarcating two controlled sensing regions.

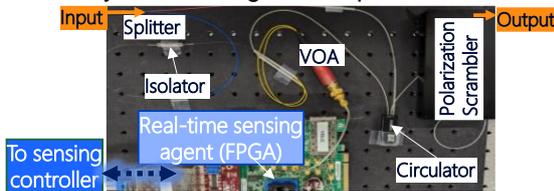

**Fig. 3:** Fiber sensing control device implementation picture.

### Real-time Sensing Control Plane

As the network itself becomes a sensor and more sensors are distributed across the network, it becomes a more complex system to manage. We implement hard slicing between communication and sensing control planes to adapt to the needs of each application and to operate the two systems separately. Sensing regions and network segments/domains do not need to overlap as they can be driven by different constraints and needs. The real-time nature of the sensing control system is critical because any delay in the reaction time can be exploited by an attacker to extract information from the optical network. As shown in Fig. 1, three sensing control layers are defined: the real-time **sensing agent** integrated in the FSCD, makes local decisions in less than 50 ns. The **sensing region controller** manages FSCDs located in a single sensing region (SR). The **multi-region sensing manager** enables fibre sensing coordination over multiple sensing regions. Each of these control layers has different control scope and reaction time. The two upper sensing layers are subject to additional propagation delay if not collocated with the FSCD. The FSCD is an instrument of the real-time sensing control plane and is the focus of this work. Other sensing elements can be introduced, for example, to route sensing signals through an optical network, and these devices could also be controlled by the sensing control plane.

### Device Experimental Evaluation

We have implemented a real-time sensing control plane similar to the one presented for real-time edge-cloud network control [9], with a real-time FSCD local controller (agent), a sensing network segment controller and the multi-region sensing manager. The FSCD implementation, as shown in Fig. 3, includes a real-time agent implemented in an FPGA and the optical components include a 3-dB coupler, an optical circulator, an optical isolator, and the backscatter path optical gate. Each FSCD implemented in our experimental setup uses different optical gate technologies.

The FSCD real-time control agent controls SoPS and BLS independently. It has four possible states. In the first one, all sensing is enabled, in the second and third, only SoPS or BLS is enabled, respectively. In the fourth state, no sensing is enabled.

**a) Backscatter Sensing Control:** we tested the BLS control using a commercially available OTDR operating at 1550 nm and an optical fibre of ~12.8 km as shown in Fig. 4. When BLS is enabled, we detect the length of the fibre, the existence of a connector at 1.1 km and the end of the segment as shown in Fig. 5a. This figure also shows the effects of the attenuation set at the variable optical attenuator (VOA) used as the BLS gate, decreasing the return optical power down to the noise floor level. The high responsivity of the FSCD agent combined with a fast electro-optical switch used as a BLS gate allows the "obscuring" of specific sections of the fibre infrastructure by disabling the backscatter path during a configurable time interval after each detected OTDR pulse. Both the obscuring duration and its delay after each OTDR pulse are configured at the agent (FPGA). Fig. 5b shows two examples of section obscuring. In the first case, the connector at 1.1 km is obscured, while in the second, the fibre section between km 5 and km 6 is obscured. The section obscuring feature can be used to protect sensitive fibre sections while allowing pulsed DAS or OTDR systems to function in the rest of the fibre. The BLS control response time depends on the optical gate technology used to control the backscatter branch of the FSCD, see Fig. 2. To determine the BLS control response time, we set a policy in the local agent that detects OTDR pulses entering the FSCD with a fast photodetector (backscatter path control unit in Fig. 2). The FSCD agent is configured with a power level threshold such that when an unauthorized OTDR pulse is detected by the threshold crossing of the photodiode output, the FPGA sends a control signal that opens the optical gate to block the backscattered light. The total response time of the FSCDs we implemented for this study -from the moment the pulse is detected until backpropagating light is blocked- ranges from less than 340 ns (electro-optical switch) to about 3 ms (MEMs-based VOA). Note that there is one FSCD at each end of the fibre to control BLS from both sides.

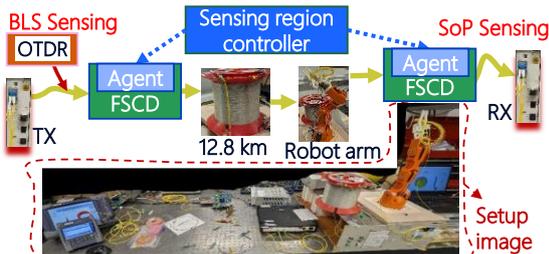

**Fig. 4:** FSCD evaluation experimental diagram.

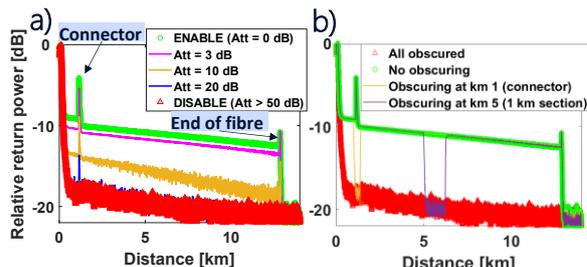

**Fig. 5:** OTDR (BLS) controlled by the fiber sensing control device, a) BLS control by VOA gate, b) Section "obscuring".

**b) State-of-polarization sensing control:** we use a polarization scrambler to obscure SoP events in a frequency range of interest. High polarization rotation rates above a certain range induce bit error rate penalty at the dual-polarization coherent receiver. We choose the scrambling rate to minimize transmission penalty while obscuring the frequency range of interest for the targeted SoP events. We use commercially available transceivers, see Fig. 4, to extract the the SoP state. The SoPS control has an activation delay below 400 µs, limited by the response time of the Adaptif A3000 SoP controller. Fig. 6 shows the effects of FSCD SoPS control with 5 ms SoP sampling period.

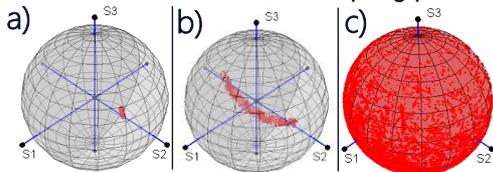

**Fig. 6:** Experimental SoP variation in Poincaré sphere when FSCD SoP sensing is: a) Enabled-no event, b) Enabled-robotic arm event, c) Disabled (event not distinguishable).

**Sensing Network Experimental Evaluation**
We emulate a two-sensing region architecture with six FSCDs, as shown in Fig. 1. SR 1 represents a premise traversed by 12.8 km of fibre and SR 2 is a transport link composed of a 40 km 7-core uncoupled fibre span from Sumitomo [10].

The first fibre sensing path traverses SR1 and SR2 and represents a path owned by a network operator with sensing enabled (all FSCD in this path are deactivated), see Fig 1. The second sensing path represents a leased transport fibre that only uses one core of the multi-core transport section (SR2). The user of the second sensing path is not authorized to perform sensing, thus, the FSCDs in this path are activated. These policies were set by the sensing manager running in a general-purpose CPU with direct connection to the SR controllers that enforce the policies through the FSCD agents. Sensing region controllers and FSCD agents are implemented in FPGAs for fast reaction. Real-time control plane links between all control layers use 10G Ethernet. We measured the real-time sensing control plane reaction time at different control layers using the fastest BLS control gate (fast electro-optical switch) in the scenario of unauthorized OTDR pulse detection. The measured response time are 340 ns at the FSCD agent level, 1.9 µs[1] at sensing region controller level, and 2 ms[1] at the multi-region sensing manager level.

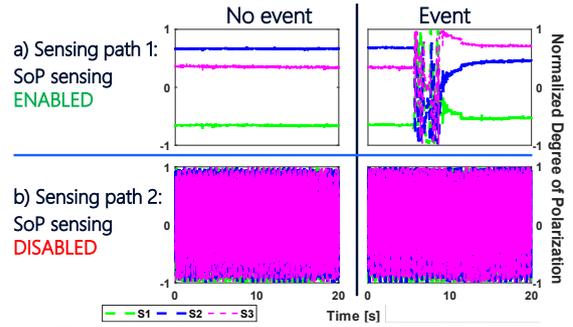

**Fig. 7:** Experimental SoP traces with and without mechanical event for: a) Sensing path 1, b) Sensing path 2.

Fig. 7 shows the measured SoP traces for the two sensing paths. We mechanically disturb the multi-core fibre and test the effectiveness of the SoPS control. In sensing path 1 (SoPS enabled), we can clearly distinguish when the fibre is disturbed from the static state, see Fig. 7a. A simple SoP slope threshold can be used to detect the disturbance. In sensing path 2, (SoPS disabled), the polarization scrambling obscures the SoP signature of the event, see Fig. 7b. The polarization scrambling frequencies of the FSCD have negligible impact on the communication signal wavelengths bit error rates.

**Conclusions**
We experimentally demonstrated the control of both backscatter- and polarization-based fibre sensing, across different fibre sensing regions using the proposed fibre sensing control device and fibre sensing control plane, providing a responsivity below 1 µs for local control. This concept can deliver high value for future optical networks as the network-as-a-sensor concept is further established, enabling control over where, when, and how fibre sensing occurs.

*(1) < 2m fiber distance between control layers in this setup. Longer distances would increase the delay in the two upper control layers (region controller and sensing manager).*

**References**

[1] T. Wild, V. Braun, and H. Viswanathan, "Joint Design of Communication and Sensing for Beyond 5G and 6G Systems," in *IEEE Access*, vol. 9, pp. 30845-30857, 2021, DOI: 10.1109/ACCESS.2021.3059488.

[2] E. Ip, Y. Huang, M. Huang, M. Salemi, Y. Li, T. Wang, Y. Aono, G. A. Wellbrock, and T. J. Xia, "Distributed Fiber Sensor Network using Telecom Cables as Sensing Media: Applications," in *Proceedings Optical Fiber Communication Conference (OFC)* 2021, pp. 1-3 DOI: 10.1364/OFC.2021.Tu6F.2

[3] C. Dorize, S. Guerrier, E. Awwad and J. Renaudier, "Capturing Acoustic Speech Signals with Coherent MIMO Phase-OTDR," ," in *Proceedings European Conference on Optical Communications (ECOC)*, Brussels, Belgium, 2020, pp. 1-4, DOI: 10.1109/ECOC48923.2020.9333283.

[4] G. A. Wellbrock, T.J. Xia, M-F Huang, S. Han, Y. Chen, T. Wang and Y. Aono, "Explore Benefits of Distributed Fiber Optic Sensing for Optical Network Service Providers," in *Journal of Lightwave Technology*, vol. 41, no. 12, pp. 3758-3766, June 2023, DOI: 10.1109/JLT.2023.3263795.

[5] S. Guerrier, A. Mecozzi, C. Dorize, C. Antonelli, L. Dallachiesa, H. Mardoyan,E. Awwad, D. Orsuti, L. Palmieri, M. Mazur,T. Hayashi, R. Ryf, and J. Renaudier, "Field Trial of High-Resolution Distributed Fiber Sensing over Multicore Fiber in Metropolitan Area with Construction Work Detection using Advanced MIMO-DAS," in *Proceedings Optical Fiber Communications Conference and Exhibition (OFC)*, San Diego, CA, USA, 2023, pp. 1-3, DOI: 10.1364/OFC.2023.W1J.5.

[6] J. E. Simsarian and P. J. Winzer, "Shake before break: Per-span fiber sensing with in-line polarization monitoring," in *Proceedings Optical Fiber Communications Conference and Exhibition (OFC)*, Los Angeles, CA, USA, 2017, pp. 1-3, DOI: 10.1364/OFC.2017.M2E.6

[7] F. Boitier, J. Pesic, V. Lemaire, E. Dutisseuil, J. Estarán, P. Jennevé, N. Le Moing, H. Mardoyan and P. Layec, "Seamless Optical Path Restoration with Just-in-Time Resource Allocation Leveraging Machine Learning," in *Proceedings European Conference on Optical Communication (ECOC),* Rome, Italy, 2018, pp. 1-3, DOI: 10.1109/ECOC.2018.8535279.

[8] M. Mazur, D. Wallberg, L. Dallachiesa, E. Börjeson, R. Ryf, M. Bergroth, B. Josefsson, N. K. Fontaine, H. Chen, D. T. Neilson, J. Schröder, P. Larsson-Edefors, and M. Karlsson, "Field Trial of FPGA-Based Real-Time Sensing Transceiver over 524 km of Live Aerial Fiber," in *Proceedings Optical Fiber Communication Conference (OFC)* San Diego, CA, USA, 2023, pp. 1-3, DOI:10.1364/OFC.2023.Tu3G.4.

[9] M. Szczerban, N. Benzaoui, J. Estaran, H. Mardoyan, A. Ouslimani, A-E. Kasbari, S. Bigo, and Y. Pointurier, "Real-time control and management plane for edge-cloud deterministic and dynamic networks," *in Journal of Optical Communications and Networking*, vol. 12, no. 11, pp. 312-323, November 2020, DOI: 10.1364/JOCN.397020.

[10] T. Hayashi, T. Taru, O. Shimakawa, T. Sasaki and E. Sasaoka, "Low-crosstalk and low-loss multi-core fiber utilizing fiber bend," *in Proceedings Optical Fiber Communication Conference (OFC), Los Angeles, CA, USA, 2011, pp. 1-3.* DOI: 10.1364/OFC.2011.OWJ3